\documentclass[12pt,preprint]{aastex}
\usepackage{lscape}
\usepackage{multirow}
\usepackage{booktabs}
\usepackage{graphicx}
\usepackage{amsmath}
\usepackage{longtable}
\usepackage{rotating}

\newcommand{\msun}{M_\odot}

\newcommand{\mj}{M_{\rm J}}

\begin{document}
\title{An M Dwarf Companion and Its Induced Spiral Arms in the HD~100453 Protoplanetary Disk}

\shorttitle{HD~100453}

\shortauthors{Dong et al.}

\author{Ruobing Dong\altaffilmark{1,2,6}, Zhaohuan Zhu\altaffilmark{3,6}, Jeffrey Fung\altaffilmark{2,7},  Roman Rafikov\altaffilmark{4}, Eugene Chiang\altaffilmark{2}, Kevin Wagner\altaffilmark{5,8,9}}

\altaffiltext{1}{Nuclear Science Division, Lawrence Berkeley National Lab, Berkeley, CA 94720, rdong2013@berkeley.edu}
\altaffiltext{2}{Department of Astronomy, University of California at Berkeley, Berkeley, CA 94720}
\altaffiltext{3}{Princeton University, Princeton, NJ, 08544}
\altaffiltext{4}{Institute for Advanced Study, Princeton, NJ, 08540}
\altaffiltext{5}{Department of Astronomy/Steward Observatory, The University of Arizona, 933 North Cherry Avenue, Tucson, AZ 85721}
\altaffiltext{6}{NASA Hubble Fellow}
\altaffiltext{7}{NSERC Fellow}
\altaffiltext{8}{National Science Foundation Graduate Research Fellow}
\altaffiltext{9}{Earths in Other Solar Systems Team, NASA Nexus for Exoplanet System Science
}

\clearpage

\begin{abstract}

Recent VLT/SPHERE near-infrared imaging observations revealed two spiral arms with a near $m=2$ rotational symmetry in the protoplanetary disk around the $\sim$1.7~$\msun$ Herbig star HD~100453. A $\sim$0.3~$\msun$ M dwarf companion, HD~100453~B, was also identified at a projected separation of 120 AU from the primary. In this Letter, we carry out hydrodynamic and radiative transfer simulations to examine the scattered light morphology of the HD~100453 disk as perturbed by the companion on a circular and coplanar orbit.
We find that the companion truncates the disk at $\sim$45 AU in scattered light images, and excites two spiral arms in the remaining (circumprimary) disk with a near $m=2$ rotational symmetry. Both the truncated disk size and the morphology of the spirals are in excellent agreement with the SPHERE observations at $Y$, $J$, $H$, and $K1$-bands, suggesting that the M dwarf companion is indeed responsible for the observed double-spiral-arm pattern. Our model suggests that the disk is close to face on (inclination angle $\sim$5 degree), and that the entire disk-companion system rotates counterclockwise on the sky. The HD~100453 observations, along with our modeling work, demonstrate that double spiral arm patterns
in near-infrared scattered light images can
be generically produced by companions,
and support future observations to identify the companions responsible for the arms observed
in the MWC 758 and SAO 206462 systems.

\end{abstract}



\section{Introduction}\label{sec:intro}

Direct imaging observations at near-infrared (NIR) wavelengths using Subaru and the Very Large Telescope (VLT) have recently discovered double-spiral-arm patterns in three gaseous protoplanetary disks: SAO 206462 \citep{muto12, garufi13}, MWC 758 \citep{grady13, benisty15}, and HD~100453 \citep{wagner15-100453}. The primaries in all three cases are Herbig stars, with estimated masses and ages of about 1.7 $\msun$ and 8 Myr \citep[SAO 206462,][]{muller11, vanboekel05}, 1.8 $\msun$ and 3.5 Myr \citep[MWC 758,][]{chapillon08, meeus12}, and 1.7 $\msun$ and 10 Myr \citep[HD~100453,][]{meeus02, dominik03, collins09}. The double-arm patterns in these systems are global; they span from 10s to $\sim$100 AU from the primary, and extend over at least 180 deg azimuthally. The arms are rather open, with pitch angles (the angle between the azimuthal direction and the tangent of the arms) on the order of 10$^\circ$ to 20$^\circ$. Most strikingly, in all cases the two arms are in approximate $m=2$ rotational symmetry. 

Although planets were immediately recognized as the primary suspects responsible for these features, initial attempts to fit the shape of the observed spirals using linear density waves theories \citep[e.g.,][]{rafikov02} were not satisfactory \citep[e.g.,][]{benisty15}. Also, as only one arm on each side of the planet's orbit is predicted by these theories, two planets in roughly symmetric locations in the disk would be needed to account for the binarity of the arms; for this arrangement
to apply to all three disks 
seems too contrived. Another possible mechanism to excite $m=2$ arms in disks is gravitational instability, as shown in \citet{dong15-giarms}; however, to trigger the instability $M_{\rm disk}/M_\star\gtrsim0.1$ is required, while these disks are probably not massive enough \citep[e.g.,][]{andrews11}.

Very recently, \citet{dong15-spiralarms} found that a massive companion on the order of 10 $\mj$ can excite two roughly symmetric arms inside its orbit, with morphologies, contrasts, and pitch angles closely resembling the observations of SAO 206462 and MWC 758. As explored in more detail by \citet{zhu15-densitywaves} and \citet{fung15}, there are two key differences between density waves excited by massive companions in the fully nonlinear regime and by less massive companions in the linear regime: in the former case (1) a secondary arm is excited in addition to the primary arm pointing towards the companion (with the azimuthal separation being a function of the companion mass), and (2) the pitch angles of the arms are much larger.

While the predicted companions SAO 206462 b and MWC 758 b have not yet been observationally confirmed, the latest VLT/SPHERE NIR imaging of HD~100453~by \citet{wagner15-100453} provides a direct link between a companion and its induced double spiral arms: both the arms {\it and} the companion outside the arms, HD~100453~B, have been imaged. HD~100453~B was first discovered by \citet{chen06}, and confirmed later to be a co-moving M dwarf with an estimated age and mass of  $\sim$10 Myr and $\sim0.3\msun$ \citep{collins09, wagner15-100453}. It is located at a projected separation of 120 AU from the primary, and its orbit is consistent with being coplanar with the disk and circular as constrained by current data with a baseline of 12 years (meanwhile an eccentric orbit cannot be ruled out based on current data). In this Letter, we combine hydro and Monte Carlo radiative transfer (MCRT) simulations to show that HD~100453~B can indeed be responsible for the arms seen in SPHERE NIR observations. The hydro and MCRT simulations are introduced in Section~\ref{sec:setup}; the modeling results are presented in Section~\ref{sec:results}; and a summary and discussion are given in Section~\ref{sec:summary}.


\section{Hydro and MCRT Simulations}\label{sec:setup}

The hydro and MCRT simulations in this work largely follow the scheme in \citet[][see also \citealt{zhu15-densitywaves}]{dong15-spiralarms}, and are only briefly summarized here. We carry out a 3D global hydro simulation using the newly developed grid-based code Athena++ (Stone et al., in prep.) to calculate the gas density structures in a protoplanetary disk in the presence of a companion (Section~\ref{sec:hydro}). The resulting 3D density structure is read into the \citet{whitney13} 3D MCRT code to produce synthetic model images at four NIR bands ($Y$, $J$, $H$, and $K1$), assuming that the dust responsible for these scattered light images is uniformly mixed with the gas. Finally, synthetic images are convolved with Gaussian point spread functions (PSF) to achieve an angular resolution comparable to real observations. In both the hydro and the MCRT simulations, we adopt parameters for both disk and companion consistent with current observational constraints.

\subsection{Hydrodynamical Simulations}\label{sec:hydro}

The hydro grid is in spherical coordinates $r$ (radial), $\theta$ (polar), and $\phi$ (azimuthal) with uniform spacing in log $r$, $\theta$, and $\phi$. The simulation box has $168\times148\times512$ grid cells in the $r\times\theta\times\phi$ directions, and covers $12-96$ AU\footnote{We experimented with different sizes of the inner boundary in the hydro simulation, and verified that the size of the inner boundary does not affect the morphologies and contrasts of the arms; only the absolute brightness of the disk is affected
by the different degrees of shadowing.}
in $r$, 52$^\circ$ above and below the midplane in $\theta$, and the entire $2\pi$ in $\phi$. Initially the disk surface density $\Sigma_0$ falls with radius as $\Sigma_0\propto 1/r$, and the disk aspect ratio is $h/r\propto r^{0.25}$ with $h/r=0.18$ at 30 AU. The surface density normalization does not affect the hydro simulation results because self-gravity is
not included.\footnote{\label{foot:shock} Although the total disk mass is undefined in the hydro calculation and is only normalized later in the MCRT 
process, simple estimates show that angular momentum transport and the consequent accretion driven by the observed spiral shocks can cause substantial disk mass depletion on a timescale of order  $10^6$ yr. This suggests that the disk may have a lower than usual mass.} The disk density and azimuthal velocity are initialized in hydrostatic equilibrium \citep[e.g.][]{nelson13}. The equation of state is locally isothermal; thus heating due to spiral shocks is not included. A constant $\alpha$-viscosity with $\alpha=10^{-4}$ is applied. We use outflow boundary condition at the inner boundary, and all quantities are fixed at their initial values at the outer boundary. The 0.3 $\msun$ companion is fixed on a coplanar and circular orbit at 120 AU around the 1.7 $\msun$ primary. We run the simulation for 37 companion orbits. Within our integration time we verify that the disk density structure has settled into a quasi-steady state after 10 orbits; in particular, by that time, the disk has been fully truncated by the companion at $\sim$45~AU in scattered light images (about 1/3 of the distance to the companion; see Section~\ref{sec:results}). Future 3D hydro simulations with significantly longer integration time can test whether the disk has truly settled into a long-term steady state. We note that in HD~100453 the disk is not truncated as the viscous torque balances resonant torques at the locations of mean motion resonances as in planet-opened gaps, and the truncation does not take place on the viscous timescale; instead, the disk is truncated by spiral shocks, and the process takes place on the dynamical timescale, i.e., in a few orbits.

In our hydro model, the scale height of the disk is slightly higher than expected based on passively irradiated disk models (e.g., \citealt{chiang97}; $h/r\lesssim0.1$ at the same radius). The aspect ratio $h/r$ is constrained by the pitch angles and the locations of the observed arms in the SPHERE data (see Section~\ref{sec:results}). As shown in \citet{zhu15-densitywaves} and \citet{fung15}, the pitch angles of companion-induced spiral arms sensitively depend on disk scale height. A smaller scale height makes less open arms and shifts their azimuthal positions at a given radius, resulting in a worse fit to the observations. Experiments show that a scale height profile $\sim20\%$ lower than the one adopted here results in an unacceptable match between the scattered light model images with observations. One possible physical motivation for our choice of a larger scale height is spiral shock heating of the disk \citep{lyra15}. As the companion drives spiral shocks into the inner disk, the dissipation of the shock heats the local disk material, resulting in an increase in disk scale height. Future hydrodynamics calculations properly taking into account disk heating from both spiral shocks and the irradiation from the central star will address this issue more directly.

\subsection{Monte Carlo Radiative Transfer Simulations}\label{sec:mcrt}

We post-process the density distribution obtained in Section~\ref{sec:hydro} with a 3D MCRT calculation using the \citet{whitney13} code, which has been extensively used to model protoplanetary disks (e.g. \citealt{hashimoto12,zhu12,dong12pds70,follette13,grady13,dong15-gaps,wagner15-169142}). Both synthetic full intensity (FI) and polarized intensity (PI=$\sqrt{Q^2+U^2}$, where $Q$ and $U$ are two components in the Stokes vector) images\footnote{In this work, the physical quantity recorded in all model images is the specific intensity in units of [mJy~arcsec$^{-2}$], or [10$^{-26}$ ergs~s$^{-1}$~cm$^{-2}$~Hz$^{-1}$~arcsec$^{-2}$].} at $Y$, $J$, $H$, and $K1$-band are produced using 1 billion photon packets. The model is viewed assuming HD~100453 is at a distance of 114~pc \citep{perryman97}, and at a modest 5$^\circ$ inclination (i.e., nearly face on) such that the line joining the primary to the companion defines the major axis and the primary arm (see section 3) is on the side
closer to the observer. The full-resolution images from the MCRT pipeline are then convolved by a Gaussian point spread function (PSF) with a full width half maximum (FWHM) of $0.025\arcsec$ at $Y$-band, $0.03\arcsec$ at $J$-band, $0.04\arcsec$ at $H$-band, and $0.05\arcsec$ at $K1$-band. These convolving PSFs are chosen to achieve the diffraction limited angular resolution at each wavelength for an 8.2 meter VLT mirror.

The 3D disk structure in the MCRT simulation is identical to that of the hydro model. The central source is a Herbig Ae star with a temperature of $7000$~K and a radius of $2.5R_\odot$, appropriate for a 1.7 $\msun$ pre-main-sequence star 10 Myr old \citep{baraffe98}. The grains in the disk are assumed to be the standard interstellar medium (ISM) grains\footnote{Although there is evidence indicating grain growth to some extent in this system \citep{meeus01, meeus02}, our assumptions of ISM grains is adequate for the simple purpose of fitting the general morphology of the arms in narrow band scattered light images.} as in Kim et al. (1994). They are made of silicate, graphite, and amorphous carbon. The size distribution of the grains is a smooth power law in the range of 0.02-0.25~$\micron$ followed by an exponential cut off beyond 0.25~$\micron$. The optical properties of the grains can be found in Figure 2 in \citet{dong12cavity}. These grains are assumed to be dynamically well coupled to the gas as they are small enough, and thus they have a volume density linearly proportional to the gas density. The total mass of the ISM grains in our model is assumed to be $M_{\rm small\ grains}=5\times10^{-6}\msun$ (or on the order of $M_\oplus$). This corresponds to, for example, a total gas disk mass of 0.005~$\msun$, a 100:1 gas-to-dust-mass-ratio, and a $10\%$ dust mass fraction in these small ISM grains\footnote{The remaining 90\% of the dust mass is assumed to be in the grains that have grown to larger sizes and settled to the disk midplane and thus do not affect NIR scattering.}. Note that the assumed disk mass is at the lower end for protoplanetary disks, as motivated by the relatively old age of the system and the anticipated disk accretion driven by the companion
(see footnote \ref{foot:shock}). The relatively low disk mass is also consistent with \citet{collins09}, who suggested that the system may be in transition from a gas-rich protoplanetary disk to a gas poor debris disk based on CO observations, and with \citet{meeus01,meeus02}, who determined that this object could have a deficit of small grains by fitting the SED of the object. On the other hand, we note that the total mass of the small grains $M_{\rm small\ grains}$ cannot be well constrained by simply fitting the general morphology of the arms. Experiments showed that varying $M_{\rm small\ grains}$ by a factor of 5 above or below the assumed value has negligible effects on the morphology of the arms; meanwhile the absolute surface brightness of the disk weakly depends on $M_{\rm small\ grains}$: a factor of 5 increase in $M_{\rm small\ grains}$ results in $\sim25\%$ increase in the surface brightness.


\section{Modeling Results}\label{sec:results}

\begin{figure}
\begin{center}
\epsscale{0.9} \plotone{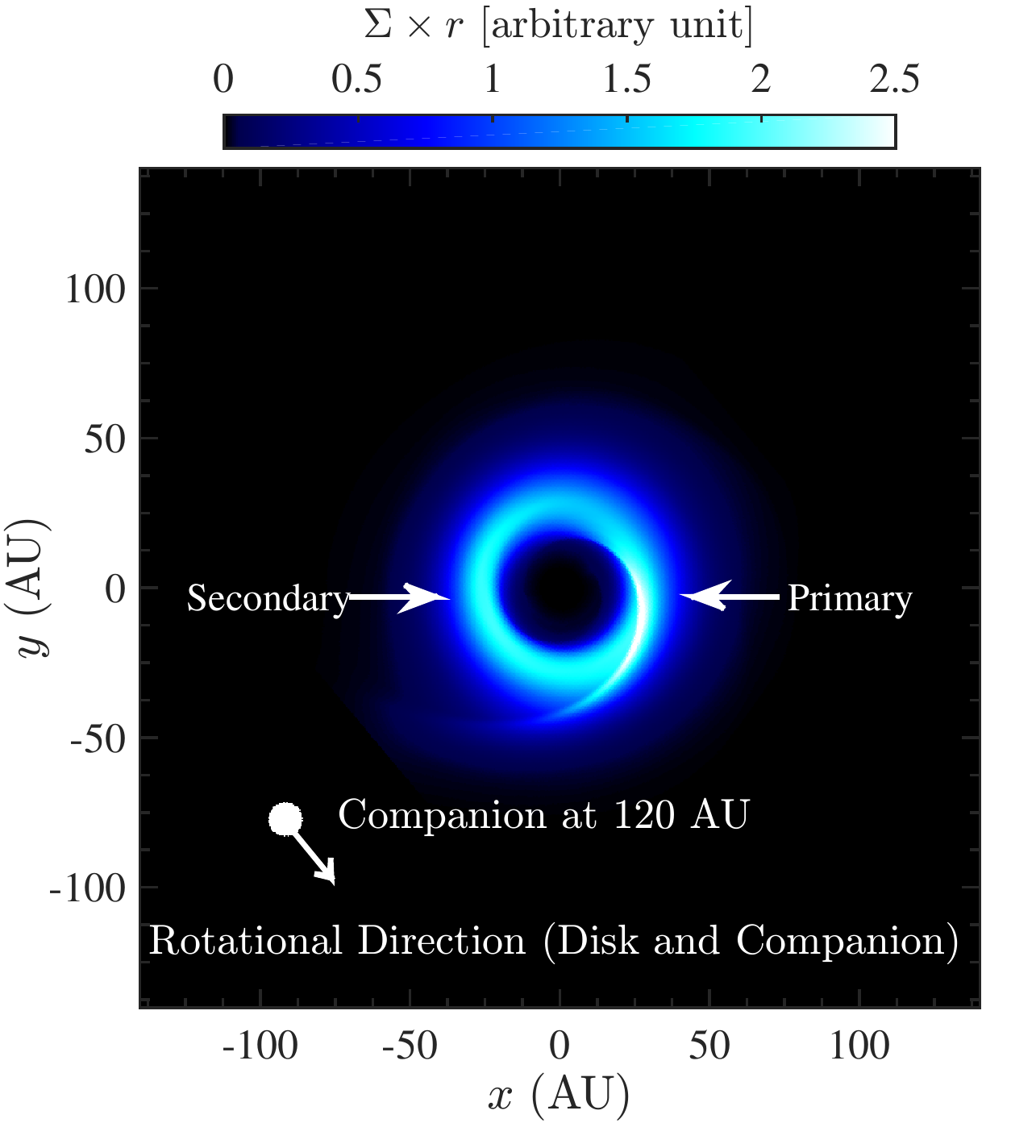}
\end{center}
\figcaption{Surface density of the hydro model (scaled by $r$ to compensate for the initial $\Sigma_0\propto1/r$ profile). The inner hole at the center marks the inner boundary in the hydro grid. The location of the companion is marked by the white dot. The companion is located at $r=$120 AU and position angle 130$^\circ$ (east of north). The gap has been fully opened, and the density structure of the disk has reached a quasi-steady state.
\label{fig:sigma}}
\end{figure}

The surface density of the hydro model is shown in Figure~\ref{fig:sigma}. The 0.3$~\msun$ companion quickly truncates the disk during the first $\sim$10 orbits, and excites two spiral arms in the remaining inner disk (the circumprimary disk). While the primary arm points roughly towards the perturber, the secondary arm is located  roughly $180^\circ$ away from the primary arm, consistent with the empirical scaling relation between the separation of the two arms and the mass of the perturber found in \citet{fung15}. At $r=20$ AU, the contrast 
$(\Sigma-\Sigma_{\rm b})/\Sigma_{\rm b}$
of the arm to the background surface density $\Sigma_{\rm b}$ is $\sim$40\% for the primary and $\sim$25\% for the secondary, while at $r=30$~AU it is $\sim$57\% for the primary and $\sim$34\% for the secondary. 

\begin{figure}
\begin{center}
\epsscale{0.9} \plotone{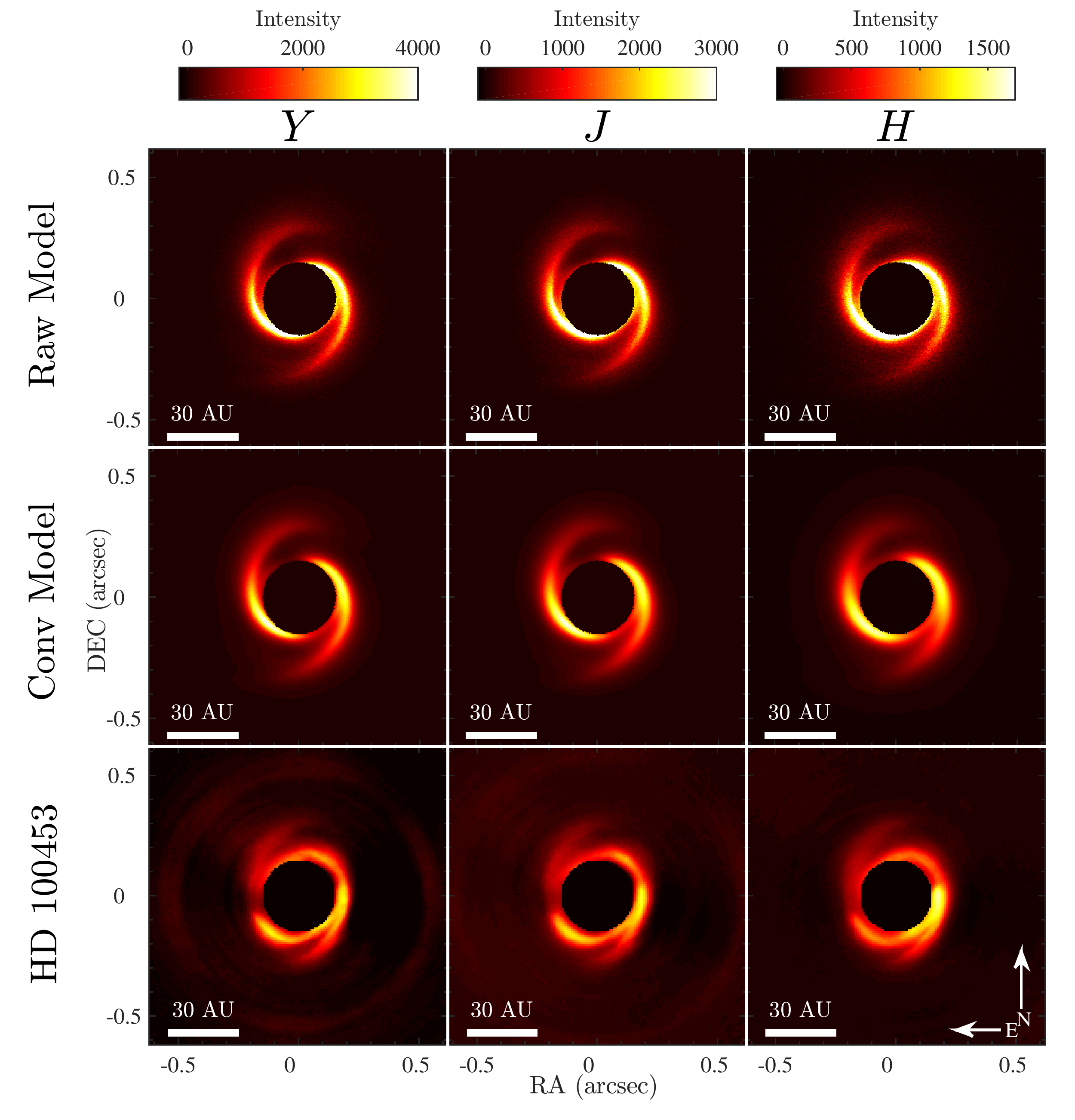}
\end{center}
\figcaption{$Y$, $J$, and $H$-band full resolution (top row) and convolved (middle row) model images, alongside observations of HD~100453 taken by the integral field spectrograph on VLT/SPHERE at the same wavelengths \citep{wagner15-100453}. The inner 0.15$\arcsec$ (17 AU), the IWA in VLT observations, is masked out in all panels. The companion (HD~100453~B) is located at position angle 130$^\circ$ (east of north) and a projected separation of 120 AU from the center and is off the scale of the plots. The units are mJy arcsec$^{-2}$ for the model images, and arbitrary for the VLT/SPHERE observations.
\label{fig:yjh}}
\end{figure}

\begin{figure}
\begin{center}
\epsscale{1} \plotone{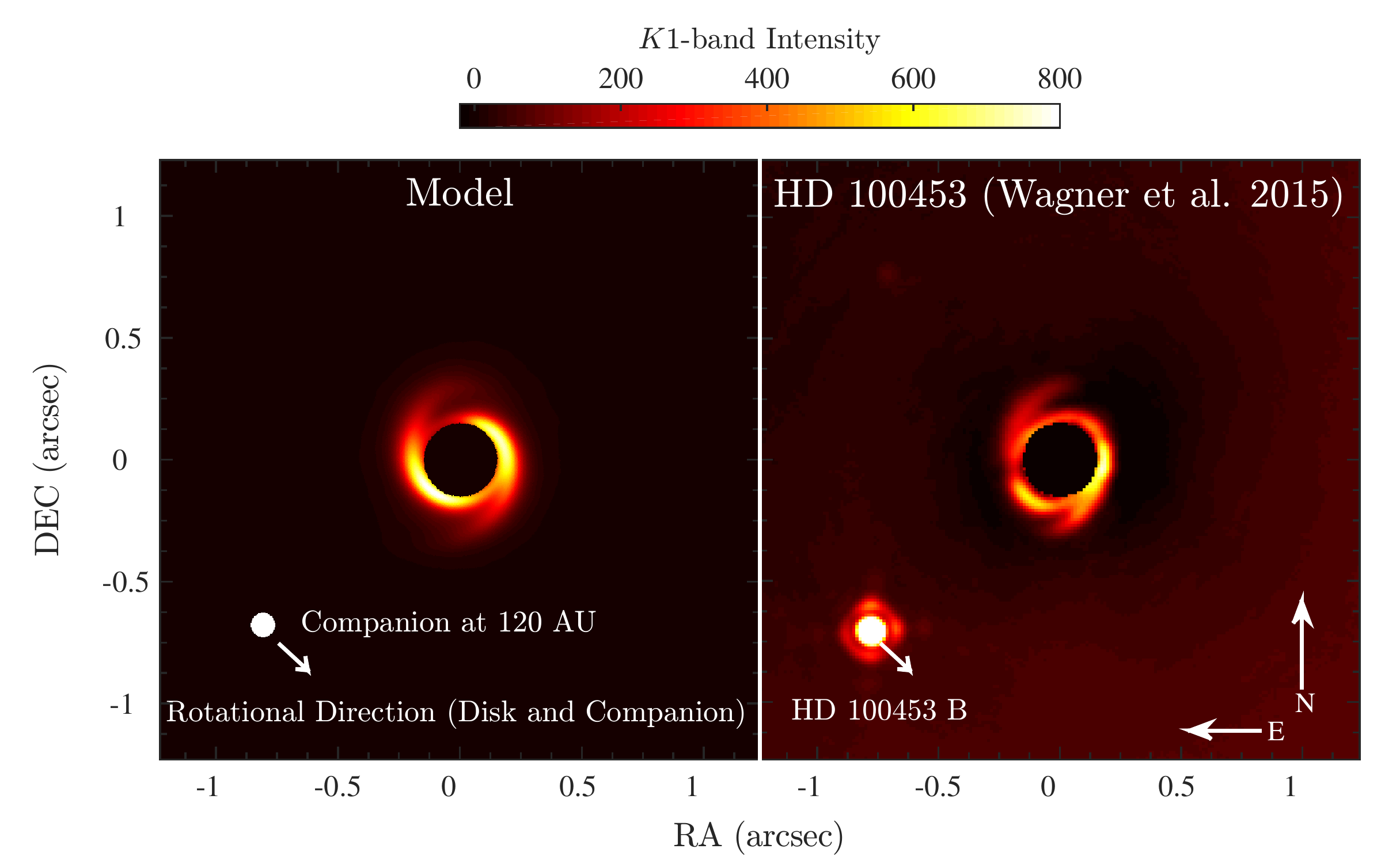}
\end{center}
\figcaption{Convolved $K1$-band model image alongside the observation of HD~100453 and HD~100453~B at $K1$-band taken by the infrared dual-band imager and spectrograph on VLT/SPHERE \citep{wagner15-100453}. HD~100453~B is at a projected separation of 120 AU from the center. The inner 0.15$\arcsec$ (IWA, 17 AU) is masked out. The units are mJy arcsec$^{-2}$ for the model image, and arbitrary for the VLT/SPHERE observation.
\label{fig:k}}
\end{figure}

\begin{figure}
\begin{center}
\epsscale{0.9} \plotone{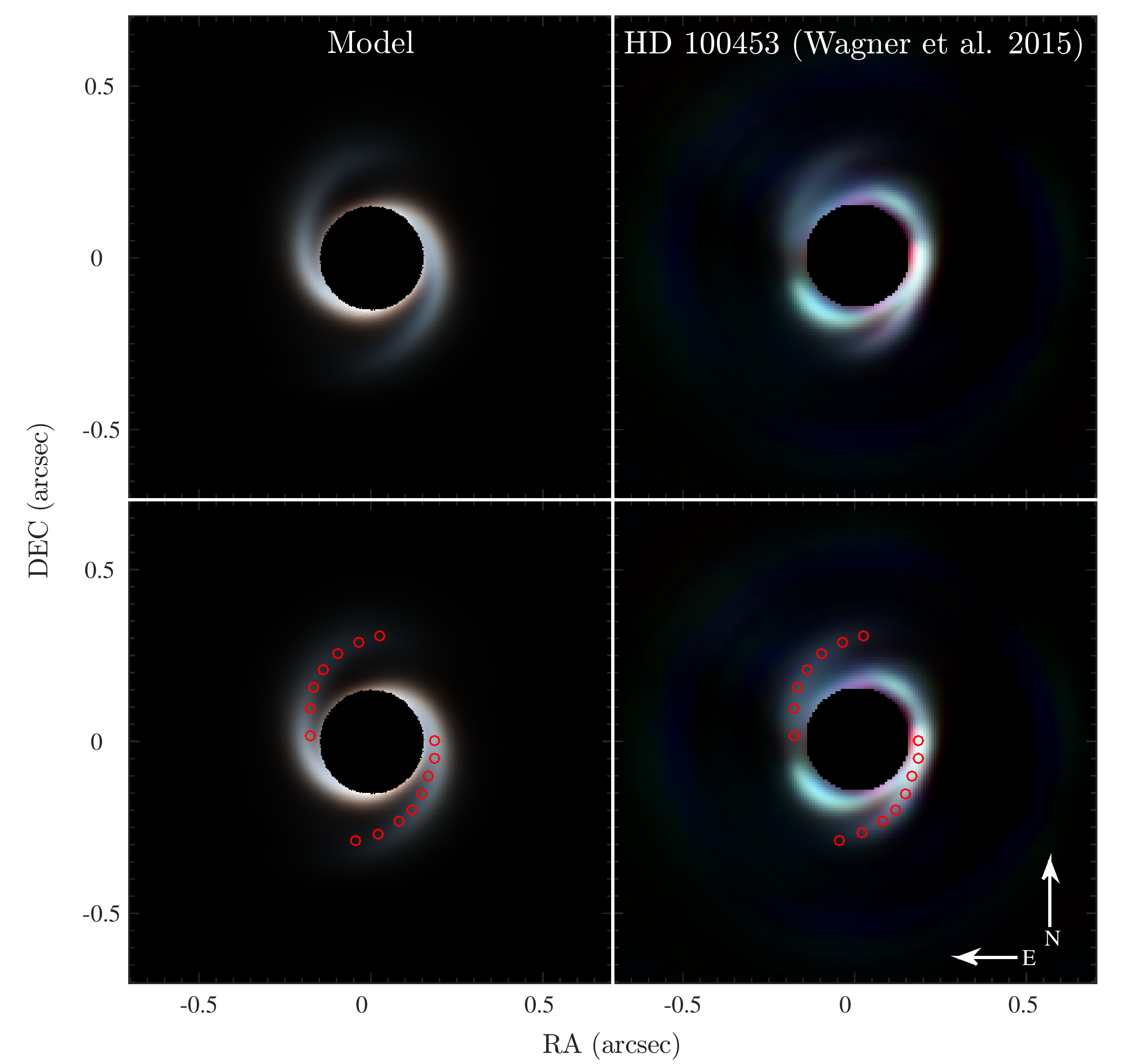}
\end{center}
\figcaption{Tri-color (RGB=HJY) images of the model and SPHERE observations of HD~100453. $Y$ and $J$-band images of both the model and the observations have been convolved to achieve the same angular resolution as the SPHERE $H$-band image. 
The three bands in both panels have been linearly stretched in the same way, with 0 (black) and 1 (bright) in each color corresponding to the minimum and maximum of the intensity in each band.
Red dots in the bottom row trace
the locus of the observed arms;
these same red dots are overlaid in the model
image for comparison. The match is good, though not perfect.
\label{fig:coord}}
\end{figure}

Full intensity model images at $Y$, $J$, $H$, and $K1$-bands are shown in Figures~\ref{fig:yjh} and \ref{fig:k} (all model images in this Letter are oriented such that the companion is at a position angle 130$^\circ$
measured east of north),
alongside the VLT/SPHERE observations of HD~1000453 at these wavelengths \citep{wagner15-100453}. The disk and the companion in our model are both in counterclockwise rotation (Figure~\ref{fig:k}), in agreement with proper motion observations of HD~100453~B (the rotation sense of the HD~100453 disk is currently unknown). The two spiral arms are clearly visible in model images at all wavelengths, and their morphologies closely resemble the observations. The arms in the models overlap with the arms in the observations, as shown in Figure~\ref{fig:coord}, demonstrating excellent agreement in the location, pitch angle, and symmetry between the two. On the other hand, our model appears to be slightly redder than the observations, which is probably due to differences in scattering properties in the small grains. We also note that, just like in the observations, the spirals in the model are only visible from $r=0.15\arcsec=17$ AU (the inner working angle, IWA, in the observations) to $r\sim 45$ AU, where the disk is truncated.

While the \citet{wagner15-100453} VLT/SPHERE observations were full intensity (FI) only, as a prediction for future polarized scattered light imaging observations we show the synthetic polarized intensity (PI) image at $H$-band and the corresponding polarization fraction (PI/FI) in Figure~\ref{fig:pi}. Overall the morphology of the arms in the PI image is similar to the FI image. The polarization fraction is slightly lower along the arms (PI/FI $\sim0.3$) than in the background disk (PI/FI $\sim0.35$). This is caused by the fact that the scattering surface is slightly higher along the arms, which results in a slightly smaller scattering angle, and thus a lower polarization fraction.

Lastly, we have repeated the Athena++ hydro simulation of the HD~100453 disk using the \citet{fung15-thesis} GPU-based hydro code PEnGUIn with the same parameters of the disk and the companion, and examined the corresponding scattered light images using MCRT calculations. We verified that PEnGUIn's results are nearly identical to the Athena++ results presented in this work, lending confidence to the robustness of our calculations.

\subsection{Disk Morphology Inside the Arms}\label{sec:innerdisk}

\begin{figure}
\begin{center}
\epsscale{1} \plotone{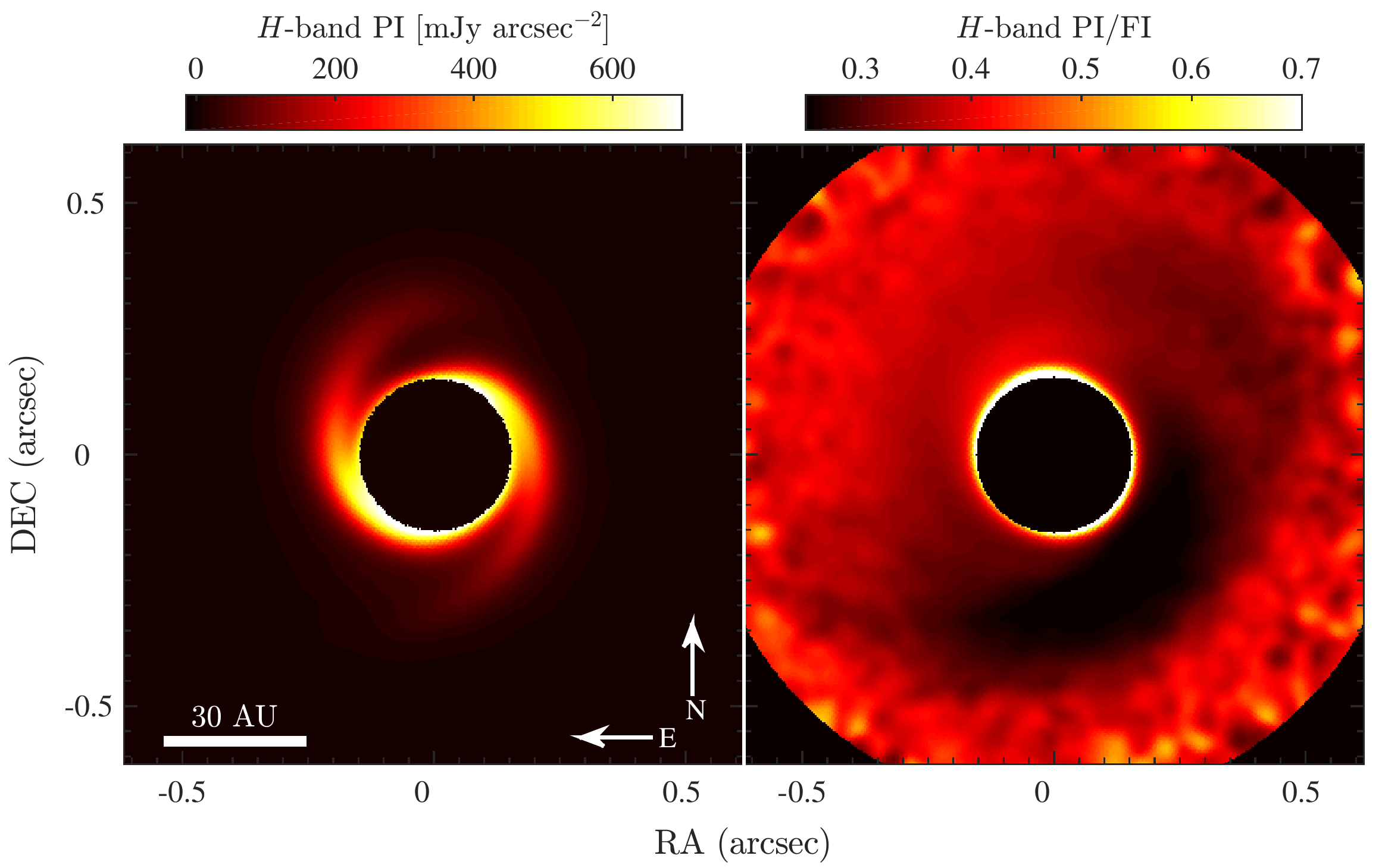}
\end{center}
\figcaption{Convolved $H$-band polarized intensity (PI) image and the polarization fraction map (PI/FI) of the model.
\label{fig:pi}}
\end{figure}

Here we comment on the observed disk morphology at 17--21 AU, inside the arms and close to the inner working angle (IWA). In this work, we focus on the morphology of the arms at $\sim$20--45 AU and do not attempt to fit the inner disk structure (the inner boundary of the disk in the hydro calculation is set to be 12 AU). In the SPHERE dataset, there appears to be a surface brightness depression just outside the IWA. \citet{wagner15-100453} interpreted this as evidence of a central gap with a wall at $\sim21$ AU, viewed at $34^\circ$ inclination. While our model cannot test this hypothesis as we have no physical inner gap at 21 AU, experiments show that the morphology of the arms induced by HD~100453~B will be dramatically altered if viewed at an inclination angle as high as $34^\circ$. Our modeling results indicate that the disk has to be nearly face on. The inclination of the disk will be examined by future gas kinematic observations using instruments such as ALMA. Another way to constrain the inclination of the system is through polarization intensity imaging. As shown in Figure~\ref{fig:pi}, the near side of the disk (the primary arm's side) has a lower polarization fraction (PF/I) than the far side of the disk. This is due to the inefficiency of the small grains in producing polarized scattered light when the scattering angle is small. The azimuthal variation in PF/I at 30 AU in our near-face-on model images is $\lesssim15\%$. A significantly higher azimuthal PF/I variation may indicate a substantial inclination of the system.

The surface brightness structure in the
``ring" at $\sim$21 AU in the observations appears asymmetric, 
and could be interpreted as follows.
First and most simply, the asymmetry may be caused by the spiral arms. As can be seen in Figures~\ref{fig:yjh} to \ref{fig:pi}, the 
regions where the arms intersect the IWA
are brighter than other portions of the
ring. We note that the southeast enhancement in our model is in good qualitative agreement with observations, while the northwest enhancement in our model is less prominent (but still clearly visible) in the observations. Other
possible causes for the observed non-axisymmetry include
a slight eccentricity in the companion's orbit; a disk warp; or shadows cast by disk structures within the IWA \citep[e.g.,][]{marino15-hd142527, dong15-shadow, pohl15}. In particular, we note that the dims at $\sim20$~AU in the SPHERE images at 2 o'clock and 9 o'clock are qualitatively similar to the dips seen on the ring in the HD~142527 system \citep{canovas13, avenhaus14, rodigas14}, which are shadows casted by the inner disk \citep{marino15-hd142527}. Lastly, the inner disk region immediately outside the inner boundary in the hydro simulation may be affected by our treatments of the inner boundary and the inner disk in the hydro simulation.


\section{Summary and Discussion}\label{sec:summary}

In this Letter, we have carried out hydrodynamical and radiative transfer simulations to examine the scattered light morphology of a protoplanetary disk around a 1.7 $\msun$ star orbited by a 0.3 $\msun$ M dwarf companion on a circular and coplanar orbit at 120 AU. We found that the companion  truncates the disk at $\sim$45 AU in scattered light images, and excites two spiral arms in the remaining (circumprimary) disk with a near $m=2$ rotational symmetry. The morphology of the spirals is in excellent agreement with the recent VLT/SPHERE observations of the HD~100453 system at $Y$, $J$, $H$, and $K1$-bands, demonstrating that the M dwarf companion HD~100453~B is almost certainly responsible for the observed double-spiral-arm pattern.

We comment now on the similarities and differences between the double-arm patterns induced by a stellar mass companion (this work) and by giant planets \citep{dong15-spiralarms, zhu15-densitywaves, fung15}. In both cases,  although the secondary arm has a surface density contrast with the background lower than the primary arm, the two arms have nearly identical surface brightnesses in scattered light images. Near-infrared surface brightness
is sensitive to grains in disk surfaces,
lifted up by the vertical motions of gas
in the arms, and is less sensitive to
gas density at the midplane (and by
extension the surface density).
In the stellar mass companion case, the two spirals span radii limited by the tidal truncation of the disk. As a result, the radial separation between the disk-perturbing companion and the spiral arms is much smaller if the companion is a giant planet. For instance, the arms in our stellar companion model are most prominent between $r\sim20$ AU and 45 AU; if the companion in our model, located at 120 AU, is replaced by a 10~$\mj$ giant planet, the arms will be most prominent at $r\sim50$ to 100 AU \citep{dong15-spiralarms, fung15}. It follows that the radial location of the arms is an ambiguous indicator of where the companion is. Nevertheless, the tip of the primary arm always points toward the companion, regardless of the companion mass.

A few predictions about the HD~100453 system can be made based on our model. The assumption that HD~100453~B is on a circular and coplanar orbit is the simplest one consistent with constraints based on 12 years of observations. A circular and coplanar orbit is able to explain the observed disk morphology, and will be further tested 
as the companion's orbit is measured at more
epochs. 
Future theoretical studies
should also assess the extent to which
the assumptions of circularity and coplanarity
can be relaxed.
Another prediction of our model is the pattern speed of the arms. As induced by the companion, the arms rotate with a pattern speed equal to
the orbital frequency of the companion, $\sim$$3^\circ$ per 10 years. This arm pattern
speed can be measured using high-resolution scattered light imaging observations in the coming years. If instead the arms are produced by some local process operating at or close to the locations of the arms (e.g., gravitational instability), the pattern speed should be
closer to the local Kepler frequency and
therefore a few times faster (e.g., the angular Keplerian velocity at 30 AU is 8 times faster than the companion's angular velocity). Finally, our model predicts that the rotational direction of the disk is counterclockwise --- the same as that observed for the companion --- and this can be tested by gas kinematic observations by ALMA.


\section*{Acknowledgments}
We thank Daniel Apai and Paul Duffell for useful discussions, and are grateful to the anonymous referee for constructive suggestions that improved the quality of the paper. This project is supported by NASA through Hubble Fellowship grants HST-HF-51333.01-A (Z.Z.) and HST-HF-51320.01-A (R.D.) awarded by the Space Telescope Science Institute, which is operated by the Association of Universities for Research in Astronomy, Inc., for NASA, under contract NAS 5-26555. E.C. acknowledges support from NASA and the NSF. J.F. is grateful for the support from NSERC and the Center for Integrative Planetary Science at the University of California, Berkeley. R.R. is an IBM Einstein Fellow at IAS; he acknowledges support from NSF, NASA, and Ambrose Monell Foundation. K. W. is supported by the National Science Foundation Graduate Research Fellowship Program under grant No. 2015209499. The results reported herein benefited from collaborations and/or information exchange within NASAÕs Nexus for Exoplanet System Science (NExSS) research coordination network sponsored by NASAÕs Science Mission Directorate. Numerical calculations were performed on the SAVIO cluster provided by the Berkeley Research Computing program, supported by the UC Berkeley Vice Chancellor for Research and the Berkeley Center for Integrative Planetary Science.


\clearpage

\end{document}